\newcommand{\bee}{\begin{equation}}
\newcommand{\ene}{\end{equation}}
\newcommand{\beea}{\begin{eqnarray}}
\newcommand{\enea}{\end{eqnarray}}

\newcommand{\bme}{\begin{multlined}}
\newcommand{\eme}{\end{multlined}}

\documentclass[10pt,pre,twocolumn,showpacs,amsmath]{revtex4}
\usepackage[pdftex]{graphicx} 
\usepackage{color}
\usepackage{wasysym}

\usepackage{txfonts}
\usepackage{pifont}
\usepackage{MnSymbol}

\definecolor{dgreen}{RGB}{30,150,70}
\definecolor{mmagenta}{RGB}{255,33,245}

\begin{document}

\title{Triple junction at the triple point resolved on the individual particle level}

\author{M.~Chaudhuri$^{1,2,3}$, E.~Allahyarov$^{3,4}$, H.~L\"owen$^{3}$, 
S.U.~Egelhaaf$^{5}$ and D.A.~Weitz$^{1,2}$}
\affiliation{$^1$Department of Physics, Harvard University, Cambridge, MA 02138, 
USA}
\affiliation{$^2$School of Engineering and Applied Sciences, Harvard University, 
Cambridge, MA 02138,USA}
\affiliation{$^3$Institute for Theoretical Physics II: Soft Matter, Heinrich Heine 
University, 40225 D\"usseldorf, Germany}
\affiliation{$^4$Theoretical Department, Joint Institute for High Temperatures, 
Russian Academy of Sciences (IVTAN), Moscow 125412, Russia}
\affiliation{$^5$Condensed Matter Physics Laboratory, Heinrich Heine University, 
40225 D\"usseldorf, Germany}

\begin{abstract}

At the triple point of a repulsive screened Coulomb system, a face-centered-cubic (fcc) 
crystal, a body-centered-cubic (bcc) crystal and a fluid phase coexist.
At their intersection, these three phases form a liquid groove, the triple junction.
Using confocal microscopy, we resolve the triple junction on a single particle level in a 
model system of charged PMMA colloids in a nonpolar solvent.
The groove is found to be extremely deep and the incommensurate solid-solid interface 
to be very broad.
Thermal fluctuations hence appear to dominate the solid-solid interface.
This indicates a very low interfacial energy.
The fcc-bcc interfacial energy is quantitatively determined based on Young's equation 
and, indeed, it is only about 1.3 times higher than the fcc-fluid interfacial energy close 
to the triple point.
\end{abstract}


\date{\today}

\maketitle

\begin{figure}
\includegraphics[width=0.62\linewidth]{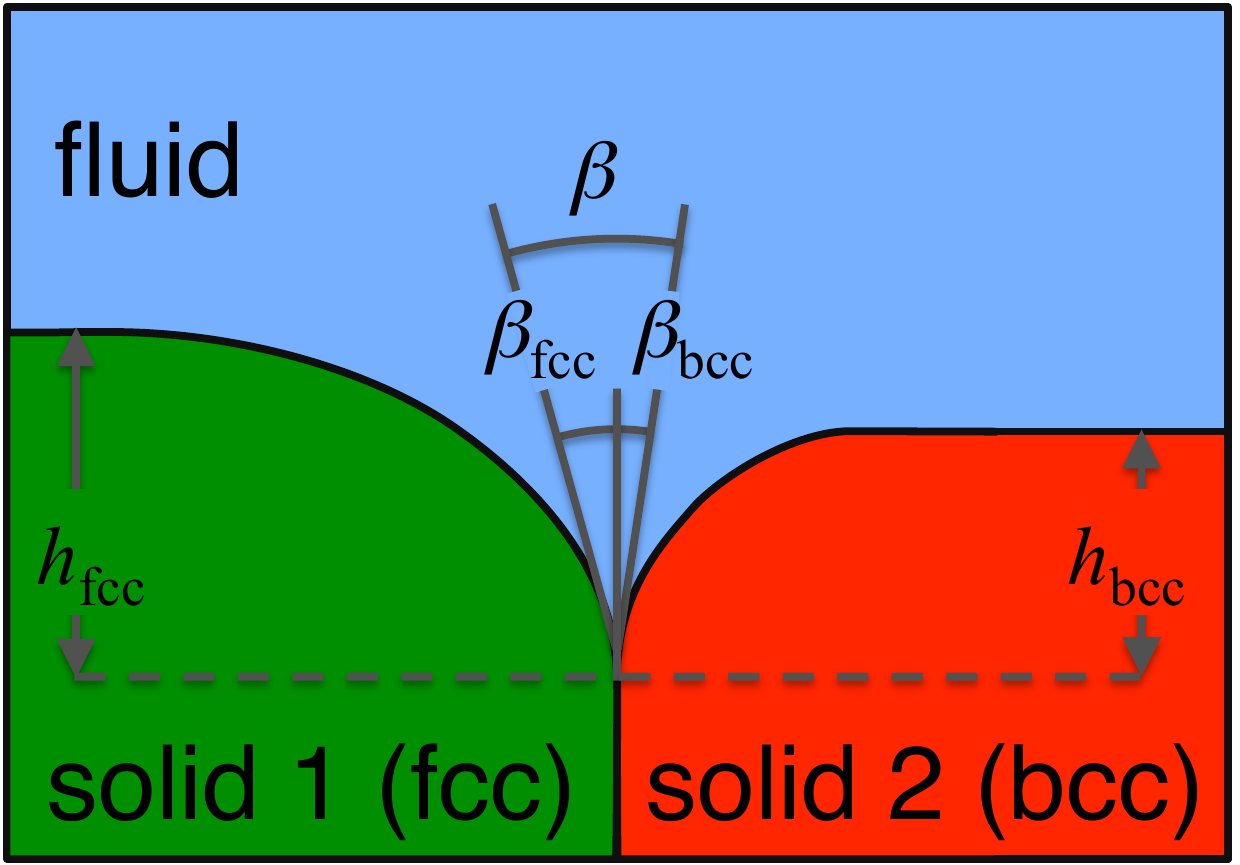}
\caption{
Schematic representation of a triple junction involving two different solids, a fcc (green) 
and a bcc (red) phase, as well as a fluid (blue).
A fluid groove can exist with a triple junction at its tip.
The dihedral angle at the tip $\beta = \beta_{\rm fcc}+\beta_{\rm bcc}$ where 
$\beta_{\rm fcc}$ and $\beta_{\rm bcc}$ are indicated.
The depths of the groove, $h_{\rm fcc}$ and $h_{\rm bcc}$, are also indicated.
}
\label{fig01} 
\end{figure}

According to the traditional Gibbs phase rule of thermodynamics~\cite{Callen_book}, 
in a 
one-component system up to three phases can coexist.
Their coexistence is represented by a triple point in the temperature-pressure phase 
diagram and a triple line in the temperature-density phase diagram.
At triple conditions, the three phases are in mutual mechanical, thermal and 
chemical equilibrium.
The three possible interfaces only occur at the same time if the interfacial energies are 
similar; if an interfacial energy is larger than the sum of the other two, this interface 
is unstable and the 
third phase intervenes.
When the three interfaces intersect, they form an interfacial line, the triple junction line 
(which is a point in the slice shown in Fig.~\ref{fig01}).
Triple junctions have been studied on the macroscopic level, for example in metals 
where liquid lenses form on top of a crystallite surrounded by coexisting vapor 
\cite{heyraud_metois_1983,lowen_ss_1990}.
Another classical example is the triple point of water where vapor, liquid water 
and ice coexist.
Such a gas-liquid-solid triple point involves two disordered and one ordered phase 
and can exist in systems governed by sufficiently long-ranged attractive 
interparticle interactions~\cite{Ilett1995,poonPRL1999,Lekkerkerker2002}.
In contrast, the coexistence of a fluid and two different solids involves two ordered 
structures and hence an interface between two crystallites that are not commensurate.
The corresponding phase behavior has 
been studied in suspensions of charged colloids \cite{Monovoukas1989, 
Sirota1989, Stevens1993, Hynninen2003, royalJCP2006, yodhPRL2010, ILMR, 
Dupont1993, Hamaguchi1997, Meijer1991}, star polymers \cite{Watzlawek1999}, 
microgels \cite{Gottwald2004} or dusty plasmas~\cite{thomas1996, morfillRMP, 
ILMR, Chu1994, fortov2005, vaulina2000}.
In these systems, a face-centered cubic (fcc) crystal, a body-centered cubic (bcc) 
crystal and a fluid can coexist.
Colloids have long been used as model systems to explore such situations, 
including nucleation~\cite{Gasser:2001aa,palbergJCP2005,
tanaka2010PNAS,Wang2012}, crystallization~\cite{Poon2009,
 oettelPRL, allahyarov2015, Zhou2011,Shabalin2016}, melting~\cite{palbergJCP, 
 yodhPRL2010, Sprakel2017}, defects \cite{Loehr2016}, glass 
 transition~\cite{Weeks:2000aa, tanaka2010NM, tanaka2012NC, Zhang2016}, 
 solid-liquid interfaces~\cite{Weeks2009, Ramsteiner2010, Schall2011, 
 Heinonen2013, Palberg2016}, solid-solid phase 
 transformations~\cite{yethiraj.PRL2004, crocker2012, yodhNat.mat.2015, 
 Maire2016} as well as the microscopic response to external 
 forces~\cite{Evers2013, Laurati2017, Sentjabrskaja2015, Schall2007}.

Using charged colloids~\cite{yethiraj_nature,yethiraj.nature}, we investigate the 
triple junction at the triple point on an individual-particle level, i.e.~including the smallest relevant length scale.
At the triple point, we find a deep and tight fluid groove between the two solid phases 
and a very broad solid-solid interface.
This indicates a small solid-solid interfacial energy and hence a considerable effect 
of thermal fluctuations.
Indeed, a quantitative determination of the interfacial energy using Young's equation 
confirms this suggestion.

The interactions of highly charged colloids in the presence of small ions 
can be described by a purely repulsive screened Coulomb (or Yukawa) effective 
pair interaction $U(r)=(Q^2/4\pi \epsilon \epsilon_0 r)\exp(-r/\lambda)$ with the 
particle charge $Q=Ze$, the elementary charge $e$, the permittivity of the vacuum 
$\epsilon_0$, the relative permittivity $\epsilon$ and the Debye screening length 
$\lambda$~\cite{Dupont1993,Hamaguchi1997,Hynninen2003}.
Phase space is completely parameterized by the Coulomb coupling parameter 
$\Gamma =Q^2/(4\pi \epsilon \epsilon_0 a \, k_{\rm B}T) = Z^2\lambda_{\rm B}/a$ 
and the screening parameter $\kappa =a/\lambda$, where $k_{\rm B}T$ is the 
thermal energy, $\lambda_{\rm B} = e^2/(4\pi \epsilon \epsilon_0 k_{\rm B}T)$ 
the Bjerrum length, $a = \rho^{-1/3}$ the mean interparticle distance and $\rho$ 
the particle number density.
For this system, using Molecular Dynamics simulations the triple 
point was located at $\Gamma  \approx 3500$ and 
$\kappa = 6.90$~\cite{Hamaguchi1997}.


\begin{figure}
\includegraphics[width=0.77\linewidth]{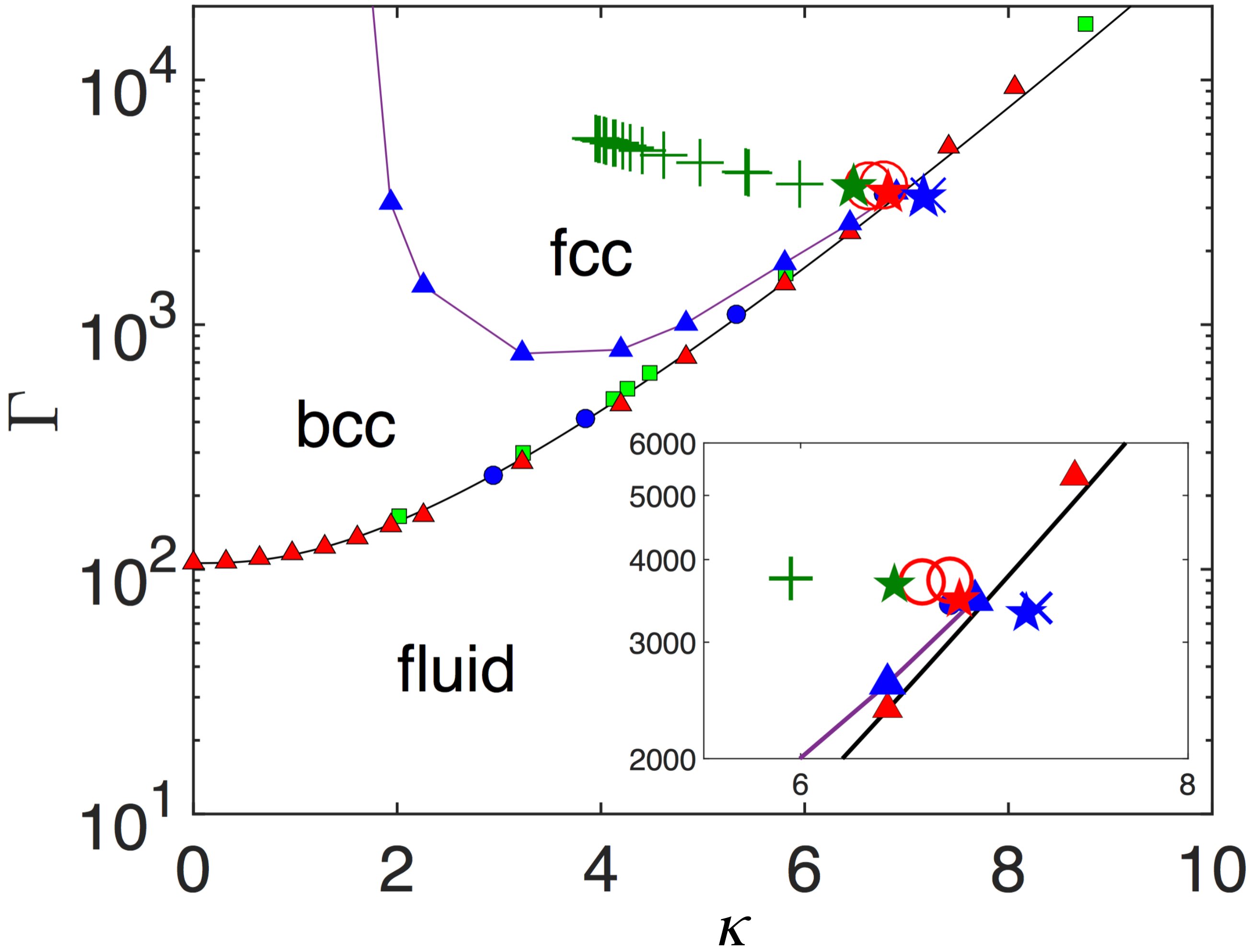}
\caption{Phase diagram of the purely repulsive Yukawa system as a function of the 
Coulomb coupling parameter $\Gamma$ and the screening parameter $\kappa$.
The experimentally investigated samples in the different single-phase regions are represented by 
{\textcolor{dgreen}{$+$}} (fcc), {\textcolor{red}{$\ocircle$}} (bcc) and 
{\textcolor{blue}{$\times$}} (fluid) and the three coexisting phases in the sample 
under triple conditions by 
{\textcolor{dgreen}{\ding{72}}} (fcc), {\textcolor{red}{\ding{72}}} (bcc) and 
{\textcolor{blue}{\ding{72}}} (fluid).
The theoretically predicted fluid-solid coexistence line is indicated by 
{\textcolor{red}{$\blacktriangle$}}~\cite{Hamaguchi1997}, 
{\textcolor{blue}{$\medbullet$}}~\cite{Meijer1991}, 
{\textcolor{green}{$\blacksquare$}}~\cite{Stevens1993} and the corresponding analytical expression 
$\Gamma = 106\,{\rm e}^{\kappa}/(1+\kappa+0.5\kappa^2)$~\cite{vaulina2000} by a solid line. 
The theoretically predicted fcc-bcc coexistence is indicated by {\textcolor{blue}{$\blacktriangle$}} 
connected by a line.
The inset shows the data close to the triple point.}
\label{fig02} 
\end{figure}
 
We use charged fluorescently-labelled poly-methyl\-meth\-acrylate (PMMA) spheres 
coated with poly-hydroxystearic acid with a radius $R \approx0.80\,\mu$m, as 
determined by dynamic light scattering.
They are suspended in a nonpolar solvent mixture of decalin ($\epsilon = 2.1$, 
density $\varrho = 0.881$~g/mL, refractive index $n = 1.48$) and tetrachloroethylene 
($\epsilon = 2.5$, $\varrho = 1.623$~g/mL, $n = 1.51$) with a ratio of 6:5 (by volume).
This mixture has a viscosity $\eta \approx 1.29$~mPa\,s as determined by rheology, a 
relative permittivity $\epsilon \approx 2.3$, a density which closely matches the particle 
density and a refractive index very similar to the one of the particles.

In nonpolar solvents charges do not readily dissociate.
However, they can be stabilized in the cores of reverse surfactant 
micelles~\cite{Hsu_2005}.
They favor the dissociation of charges from the particle surface resulting in negatively 
charged particles and charge screening by the charged reverse micelles, similar to 
the mechanism in polar solvents.
Furthermore, two neutral micelles can reversibly interact to yield two oppositely 
charged micelles; roughly one in $10^5$ micelles acquires a single elementary 
charge in this way~\cite{Hsu_2005}.
We use $20$~mM dioctyl sodium sulfosuccinate (AOT), which is well above the 
estimated critical micelle concentration of about 
$1$~mM~\cite{Hsu_2005,Mukherjee1993,Kanai2015}, and yields reverse micelles 
with an essentially concentration-independent radius 
$R_{\rm m} \approx 1.5$~nm~\cite{Hsu_2005,Zulauf1979}.
At this AOT concentration the conductivity $c \approx 80$~pS/cm, as measured using 
an immersion probe, and thus the number density of charged micelles, and hence ions, 
is estimated to be $\rho_{\rm m} = 6\pi\eta R_{\rm m} c/e^2 \approx 10^{19}$~m$^{-3}$ 
(and the number density of all, that is charged and uncharged, micelles is higher by a 
factor of about $10^5$).
This results in an estimate of the screening length 
$\lambda = 1/\sqrt{4\pi\lambda_{\rm B}\rho_{\rm m}} \approx 0.5\,\mu$m with 
$\lambda_{\rm B} \approx 24$~nm.
Electrophoretic light scattering measurements yield the normalized zeta potential 
$|e\zeta/k_{\rm B}T| \approx 3.6$. 
The particle charge number can be estimated within the DLVO theory, 
$|Z|=({\rm e}^{R/\lambda}/(1+R/\lambda))|e\zeta/k_{\rm B}T| \approx 900$ \cite{VO}.
Due to charge saturation, this is only a crude upper estimate \cite{Bocquet2002}.

We use a sample at the triple point to determine the particle charge with high 
accuracy.
A concentrated sample is prepared and less dense samples obtained by adding 
supernatant.
The sample with three coexisting phases is identified using confocal microscopy
\cite{foot}.
The three coexisting phases have slightly different particle number densities 
$\rho \approx 0.030\,\mu$m$^{-3}$ (fluid), $0.035\,\mu$m$^{-3}$ (bcc) and 
$0.040\,\mu$m$^{-3}$ (fcc, Fig.~\ref{fig02}, filled stars), which is consistent with 
the small size of the coexistence region predicted for highly charged particles 
\cite{Hynninen2003}.
At the triple point, $\Gamma \approx 3500$ and $\kappa = 6.90$ 
\cite{Hamaguchi1997}.
This suggests a screening length $\lambda \approx 0.45\,\mu$m consistent with the 
estimated $\lambda \approx 0.5\,\mu$m and a particle charge $|Q| \approx 670\,e$ 
which is below the crude upper estimate $|Q| \approx 900\,e$.
Thus, the values determined from the sample at the triple point are consistent with 
those based on the sample composition.
It also confirms that the particles are highly charged.
This can be quantified by the reduced contact value of the pair potential consisting 
of repulsive Yukawa and hard core interactions, 
$\Gamma^*= Z^2(\lambda_{\rm B}/2R)(1+R/\lambda)^{-2} \approx 900$ 
\cite{Hynninen2003,ILMR}.
Thus $\Gamma^* \gg 20$ and hard core interactions are negligible 
\cite{Hynninen2003}.
The particles hence can be treated as highly charged, point-like particles 
with purely repulsive screened Coulomb interactions.

Having determined the particle charge $Q \approx -670\,e$ and the screening 
length $\lambda \approx 0.45\,\mu$m, using these parameters we can locate 
the samples in the phase diagram (Fig.~\ref{fig02}).
The observed phase behavior is consistent with previous experimental findings 
and theoretical predictions for a repulsive screened Coulomb system 
\cite{Dupont1993, Hamaguchi1997, Hynninen2003, Meijer1991, Stevens1993, 
vaulina2000, Hamaguchi1996, Kremer1986, Sengupta1991, Smallenburg2011, 
Monovoukas1989, Sirota1989, Schoepe1998, deAnda2015}.

\begin{figure}
\includegraphics[width=0.93\linewidth]{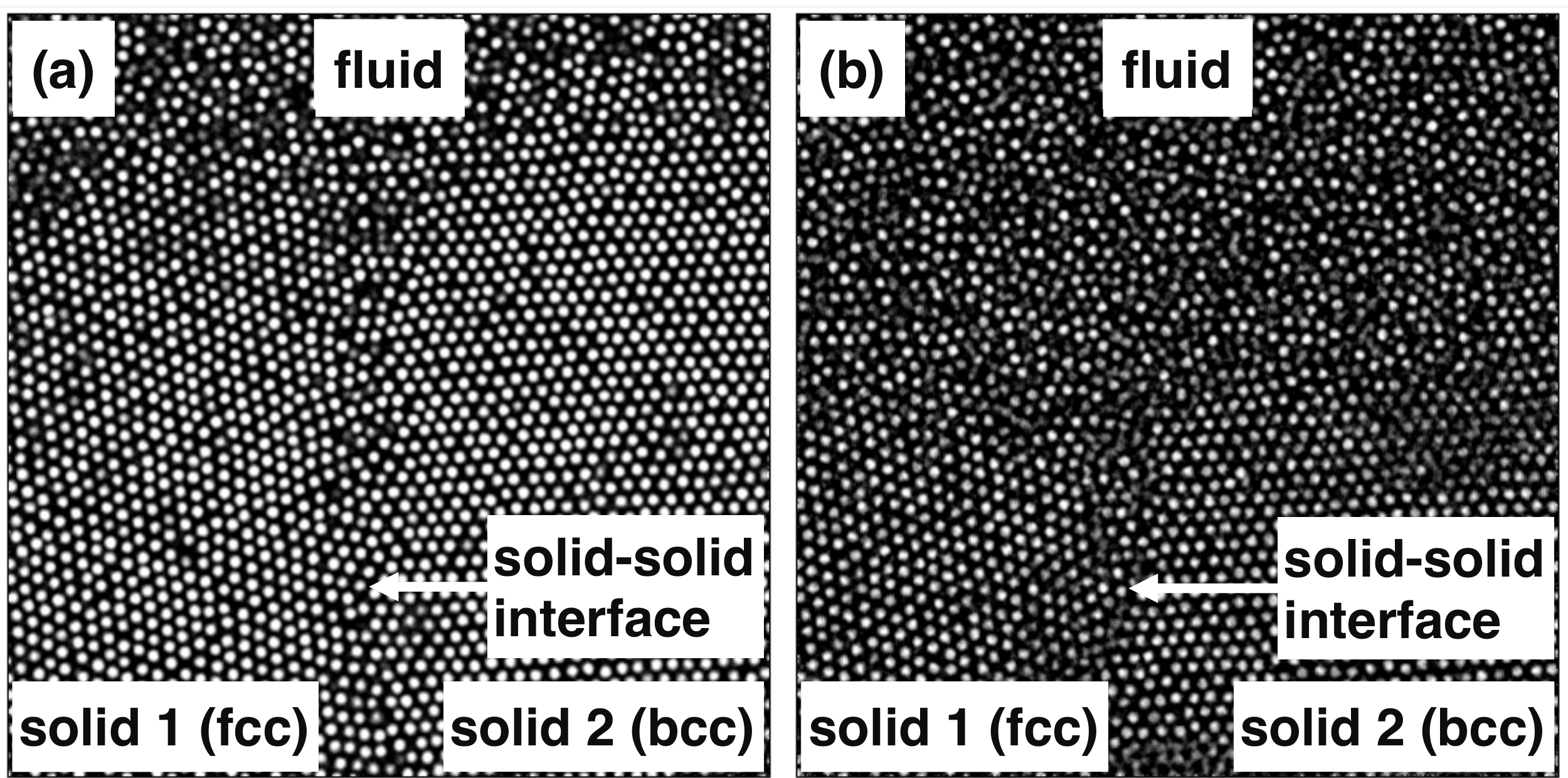}
\caption{Coexistence of two crystalline solids and one fluid as observed by confocal 
microscopy in the sample under triple conditions.
From the three-dimensional stack of images, the two outermost slices which are (a) 
$20\,\mu$m and (b) $80\,\mu$m from the cover slip, respectively, are shown.
The slices are $145 \times 145\,\mu$m$^2$.
}
\label{fig03} 
\end{figure}


The sample with three coexisting phases is investigated in more detail in the following.
Confocal microscopy images taken $20\;\mu$m from the cover slip show the coexistence of two crystalline solids and a fluid 
(Fig.~\ref{fig03}) and thus fluid-solid interfaces as well as a solid-solid interface.
At the solid-solid-fluid triple line a fluid groove starts which is many crystal layers deep.
It has a small dihedral angle that appears slightly asymmetric, reflecting the two 
different crystalline solids.
This is different in grain-boundary grooves that are formed between crystallites of the 
same structure, which have experimentally been observed in hard sphere 
systems \cite{rogers,Maire2016}.


\begin{figure}
\includegraphics[width=0.9\linewidth]{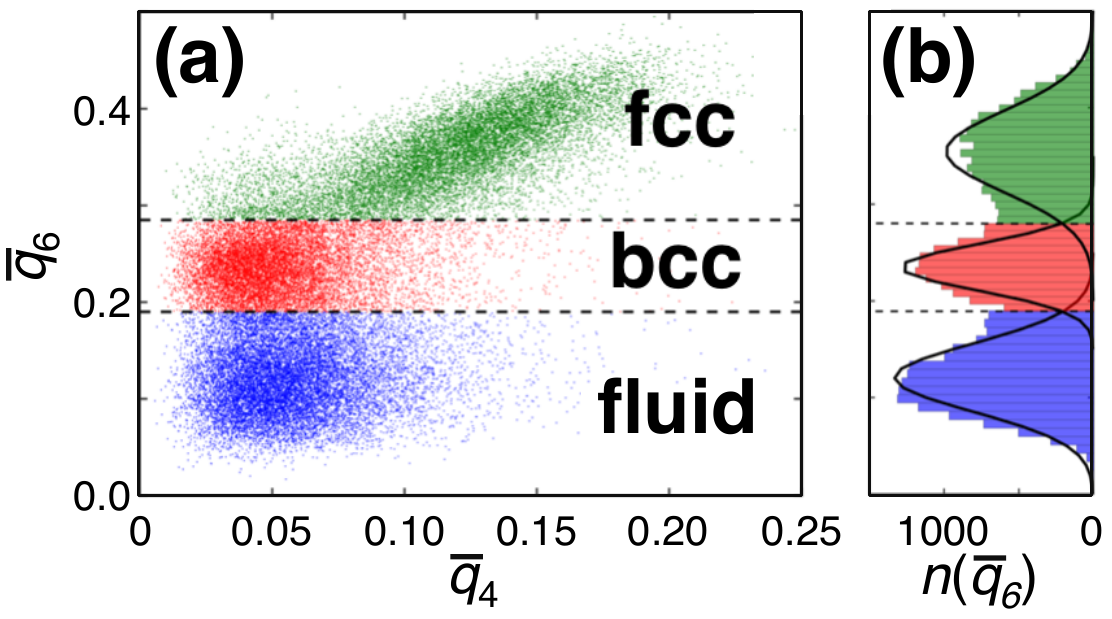}
\caption{(a) Local bond order parameters $\bar{q}_4$ and $\bar{q}_6$ of each particle 
represented by a point in the $\bar{q}_4$-$\bar{q}_6$-plane and
(b) frequency of the $\bar{q}_6$ values.
The colors indicate the different particle properties; fcc (green), bcc (red) and fluid (blue).
}
\label{fig04one} 
\end{figure}

\begin{figure}
\includegraphics[width=0.96\linewidth]{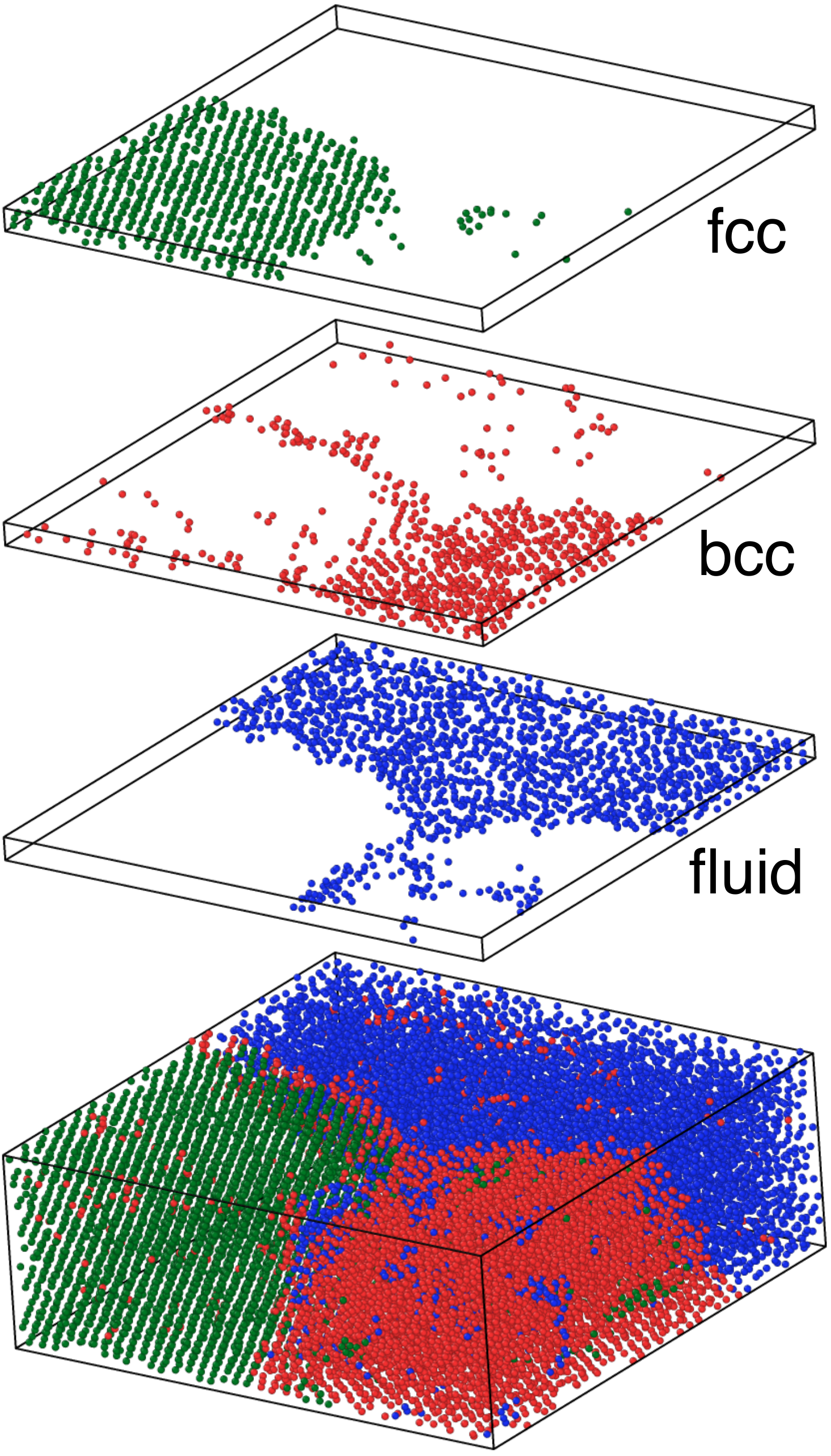}
\caption{(bottom) Slab of the sample ($125 \times 125 \times 45\,\mu$m$^3$) 
as well as (top) the individual populations in a slice of the slab 
($125 \times 125 \times 5.4\,\mu$m$^3$) with each particle represented by a 
point whose color indicates the structure of the particle; fcc (green), bcc (red) 
and fluid (blue)}
\label{fig04two} 
\end{figure}

Images of the groove are quantitatively analyzed to retrieve a profile of the groove and to 
determine the crystal structures.
This requires to determine whether a particle belongs to the fluid or one of the solids.
The particle locations are extracted from the image stacks by standard 
algorithms~\cite{crocker.grier.1996} and
for each particle the local bond order parameters $\bar{q}_4$ and 
$\bar{q}_6$~\cite{frenkel1996,allahyarov2015} are calculated (Fig.~\ref{fig04one}). 
The distribution of $\bar{q}_6$ and $\bar{q}_4$ values indicate three populations 
which can also be identified in the histogram of $\bar{q}_6$ (Fig.~\ref{fig04one}, right).
The $\bar{q}_6$ values hence can be used to guide the classification of the 
particles~\cite{frenkel1996}.
Particles with $\bar{q}_6 \leq 0.19$ are likely to belong to the fluid, particles with 
$0.19 < \bar{q}_6 < 0.28$ to the bcc crystal and particles with $\bar{q}_6 \geq 0.28$ 
to the fcc crystal.
A similar assignment is obtained based on the number of neighbors instead of the local bond order parameter $\bar{q}_6$.
If the particles are labelled accordingly, the groove and interface separating the fcc 
and bcc crystals are clearly visible (Fig.~\ref{fig04two}, bottom).
Furthermore, the orientations of the fcc and bcc bulk crystals can be determined; 
in both cases the (111) plane is oriented horizontally.
This analysis also confirms the presence of the triple junction at the intersection of the fcc-fluid, 
bcc-fluid and fcc-bcc interfaces (Figs.~\ref{fig03},~\ref{fig04two}).
While the three phases can be distinguished on a mesoscopic level, this is not the 
case on a microscopic level; individual particles are observed to spread into the 
neighboring regions (Fig.~\ref{fig04two}, top).
Fluid particles significantly penetrate into the two crystalline regions, in particular 
into the interfaces and the bcc crystal.
Moreover, individual particles with bcc structure are found in the fcc and fluid regions.
In contrast, the fcc particles are essentially confined to the fcc crystal with only very 
few particles with an fcc-like structure inside the bcc crystal.
This distribution of particles is attributed to defects and fluctuations
and partially could be due to the ambiguity in the link between crystal structure and 
$\bar{q}_6$ value, but also the difficulty to assign a crystal symmetry to an individual 
particle at finite temperature.
Furthermore, the fcc-fluid interface shows a high degree of bcc-like ordering, as 
predicted by simulations \cite{tenWolde1995}.

\begin{figure}
\includegraphics[width=\linewidth]{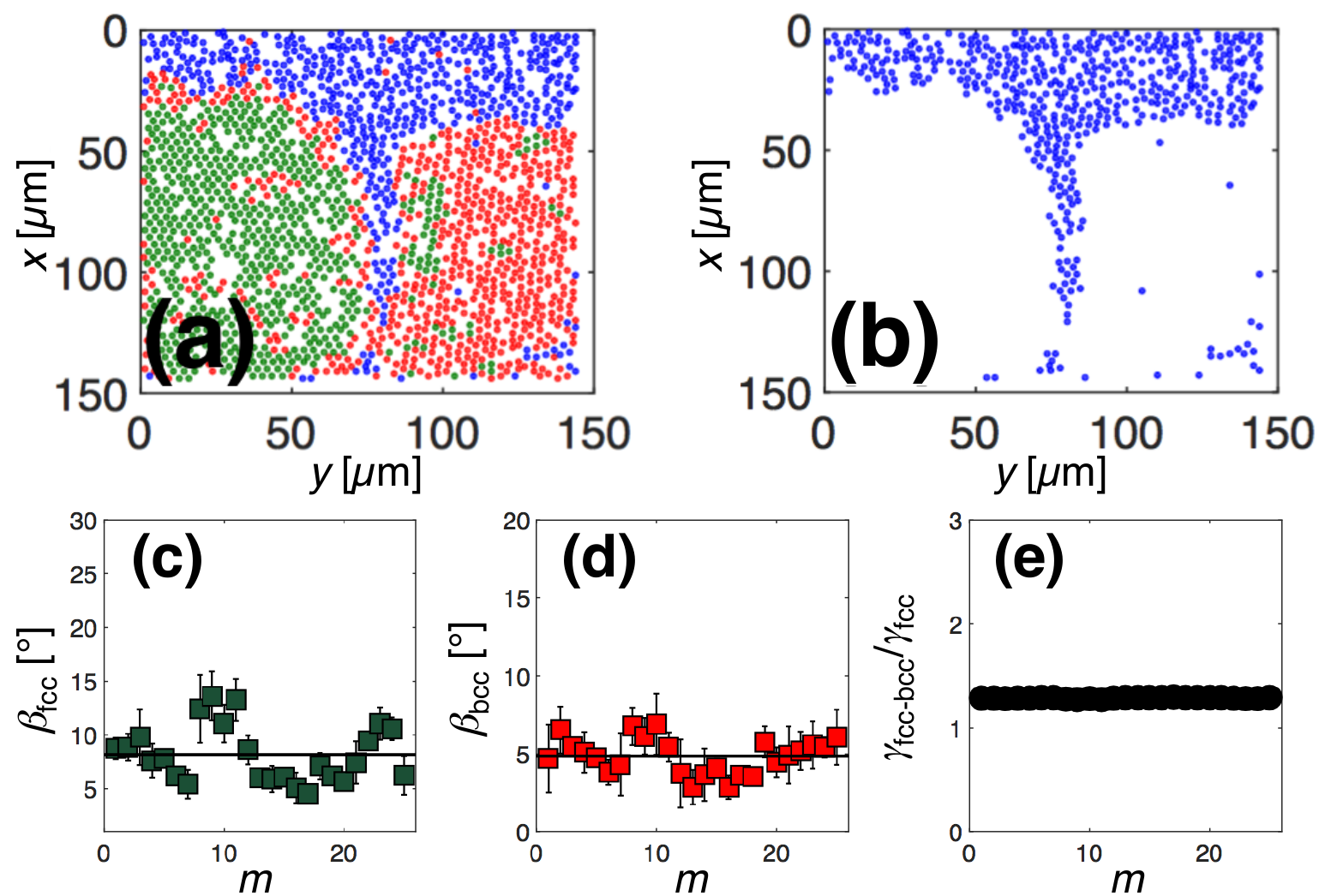}
\caption{(a) Particle layer with the colors of the particles indicating the 
corresponding phase; fcc (green), bcc (red) and fluid (blue).
(b) Only the fluid particles in the same layer.
(c) Angles of the groove towards the fcc crystal, $\beta_{\rm fcc}$, and
(d) towards the bcc crystal, $\beta_{\rm bcc}$ (Fig.~\ref{fig01}b), as well as the
(e) fcc-bcc interfacial energy $\gamma_{\rm fc}$, all as a function of the particle layer 
$m$.
}
\label{fig05} 
\end{figure}

To determine the solid-solid, i.e.~fcc-bcc, interfacial energy $\gamma_{\rm fcc-bcc}$, 
the groove is quantitatively analyzed.
The rendered three-dimensional stack is divided into $25$ quasi-two-dimensional 
$x$-$y$ planes with a thickness of about $2.2\,\mu$m (Fig.~\ref{fig05}a).
These planes are essentially parallel to the crystal planes and represent one particle 
layer (Fig.~\ref{fig04two}, bottom).
They show grooves with slightly different depths (Fig.~\ref{fig05}b).
The tip of a groove, i.e.~the triple junction, is defined as the particle that is most distant 
from the bulk fluid but still connected to the bulk fluid through other fluid particles.
The structural parameters of the grooves (Fig.~\ref{fig01}b) are quantitatively determined 
for each plane (Fig.~\ref{fig05}c,d). 
The grooves are slightly more than $100\,\mu$m wide and about $80\,\mu$m deep, which 
corresponds to about $50$ particle diameters or $35$ interlayer spacings.
Towards the fcc crystal the depth of the groove, on average $h_{\rm fcc} \approx 84\,\mu$m, 
is slightly larger than towards the bcc crystal, on average $h_{\rm bcc} \approx 72\,\mu$m.
Similar is found for the angles; $\beta_{\rm fcc} \approx 8.2^\circ$ is slightly larger than 
$\beta_{\rm bcc} \approx 4.9^\circ$, but both are very small. This results in a small 
dihedral angle $\beta = \beta_{\rm bcc} + \beta_{\rm fcc} \approx 13^\circ$.
This is in contrast to findings for grain boundaries between crystallites with the same 
symmetry where the dihedral angles are larger, about $100^\circ$~\cite{Maire2016}.

Mechanical stability of the triple junction line requires Young's condition to 
hold~\cite{Mullins1957,ratke,rogers,lin2016}. It links the interfacial free energies, $\gamma$, 
to the interface intersection angles $\beta$ (Fig.~\ref{fig01}): 
\begin{equation}
\gamma_{{\rm fcc-bcc}} = \gamma_{\rm fcc} \cos(\beta_{\rm fcc})  +  \gamma_{\rm bcc} 
 \cos(\beta_{\rm bcc})  \;\; ,
\label{eq:gamma}
\end{equation}
where the subscripts fcc, bcc and fcc-bcc refer to the fcc-fluid, bcc-fluid and fcc-bcc 
interfaces with $\gamma_{\rm bcc} \approx 0.12\,k_{\rm B}T/a^2$ and 
$\gamma_{\rm fcc} \approx 0.40\,k_{\rm B}T/a^2$~\cite{Heinonen2013} with an extended comparison of experimental and theoretical values provided by~\cite{Palberg2016}.
The angles refer to the directions of the interfaces close to the triple junction.
This equation is based on the assumption that the interfacial energy is isotropic.
It was shown to adequately describe the fluid-bcc interfaces in Yukawa 
systems close to the triple conditions, where the interfacial energy is almost 
isotropic~\cite{Heinonen2013}.
In addition, the fluid-fcc interface energy in hard sphere systems has been found to show only 
a modest dependence on the orientation~\cite{Haertel2012}.
Since the fcc-bcc interface has a fluid-like structure with many defects, it appears reasonable 
to assume that its interfacial energy is not very anisotropic.

Using Young's equation (Eq.~\ref{eq:gamma}), the fcc-bcc, i.e.~solid-solid, interfacial 
energy $\gamma_{\rm fcc-bcc}$ can be calculated.
We find $\gamma_{\rm fcc-bcc} \approx 0.52\,k_{\rm B}T/a^2 = 1.3 \, \gamma_{\rm fcc}$ 
(Fig.~\ref{fig05}e).
Thus, the interfacial energy is very small, although it involves two solid phases.
The small interfacial energy is consistent with the pronounced fluctuations observed at 
the fcc-bcc interface (Fig.~\ref{fig04two}); the interface is several layers wide and 
hence astonishingly broad.
The value of the interfacial energy is considered to be reliable as it shows only a very 
weak sensitivity to uncertainties in the angles $\beta_{\rm fcc}$ and $\beta_{\rm bcc}$ 
due to the small size of the angles and the dependence of $\gamma_{\rm fcc-bcc}$ on 
the cosine of the angles, which is close to $1$.
Thus, the uncertainty in $\gamma_{\rm fcc-bcc}$ is mainly related to the definition of the 
tip of the groove and less to the uncertainties in the determination of the angles 
$\beta_{\rm fcc}$ and $\beta_{\rm bcc}$.
Nevertheless, also the definition of the tip is not crucial, as argued in the following.

The observed coexistence of the three phases (Fig.~\ref{fig03}) implies that none of 
the three interfacial energies dominates.
For example, the sum of the two interfacial energies, $\gamma_{\rm fcc}$ and 
$\gamma_{\rm bcc}$, represents the maximum value of $\gamma_{\rm fcc-bcc}$ 
that is still compatible with the existence of a fcc-bcc interface.
For larger values, the fcc-bcc interface would be unstable towards an 
intervening fluid phase, i.e.~complete wetting.
Therefore, $\gamma_{\rm fcc-bcc} < \gamma_{\rm fcc} + \gamma_{\rm bcc}$ and, 
following the corresponding argument, 
$\gamma_{{\rm fcc}} < \gamma_{\rm fcc-bcc} + \gamma_{\rm bcc}$.
This implies 
$\gamma_{\rm fcc} - \gamma_{\rm bcc} < \gamma_{\rm fcc-bcc} < \gamma_{\rm fcc} + 
\gamma_{\rm bcc}$ and hence $0.7 \lesssim \gamma_{\rm fcc-bcc}/\gamma_{\rm fcc} 
\lesssim 1.3$.
This range is so narrow because the bcc-fluid interfacial energy is small, 
$\gamma_{\rm bcc} \approx 0.3\,\gamma_{\rm fcc}$~\cite{Heinonen2013}, which has 
been attributed to the fact that the bcc structure is relatively close to the fluid 
structure~\cite{Heinonen2013}.
This narrow range of values for $\gamma_{\rm fcc-bcc}$ hence can be 
established only based on the observation of the coexistence of the three 
phases.
Moreover, the observation of a tight groove (Fig.~\ref{fig03}) implies small angles 
$\beta_{\rm fcc}$ and $\beta_{\rm bcc}$ whose cosines are about 1.
Thus $\gamma_{\rm fcc-bcc}$ is almost the sum of the two solid-fluid interfacial 
energies, 
$\gamma_{\rm fcc-bcc} \approx \gamma_{\rm fcc} + \gamma_{\rm bcc} \approx 
1.3\, \gamma_{\rm fcc}$ (Eq.~\ref{eq:gamma}).
The value of $\gamma_{\rm fcc-bcc}$ hence is expected towards the upper limit of 
the above range of values.
This semi-quantitative argument is based on the observation of individual particles.
However, it does not require to define the tip of the groove, the observation of a tight 
groove is sufficient.
Thus, there is qualitative support as well as semi-quantitative and quantitative 
evidence for a very small fcc-bcc interfacial energy.


To conclude, we investigated suspensions of charged colloids under triple conditions, 
where a fcc crystal, a bcc crystal and a fluid coexist.
The fcc-bcc interfacial energy was found to be about 1.3 times higher than the fcc-fluid 
interfacial energy close to the triple point with 
$\gamma_{\rm fcc-bcc} \approx 0.52 \, k_{\rm B}T/a^2$.
Thus, the fcc-bcc interfacial energy is very small, despite the fact that two solid 
phases are involved.
This is consistent with the observation of broad interfaces and indicates the 
importance of thermal fluctuations also for solid-solid interfaces.
Our quantitative findings and qualitative arguments suggest that a small solid-solid 
interfacial energy not only occurs in systems with charged particles but also in 
other systems with soft interactions exhibiting a triple point.
Furthermore, also in atomic or molecular systems, e.g.~metals \cite{Hillert, Toth2010, 
Emmerich2007, Sosso2016}, similar values might be found if expressed in dimensionless units. 
Our finding hence might in general apply to triple points involving a fluid and two 
solids.
They might also be extended to more complex conditions, such as the presence of shear or 
other external fields \cite{Dhont1996, Lettinga2004, Bechinger2001}.


{\it Acknowledgments.}
We thank A.G.~Yodh and T.~Palberg for very useful discussions and  P. Ma{\ss}hoff for help 
with the figures.
M.C.~is supported by a Marie-Curie international outgoing fellowship within the 
European Union's $7^{\rm th}$ Framework Program (Grant No.~327168).
We acknowledge support by the Deutsche Forschungsgemeinschaft (Grants 
LO418/19-1, EG268/6-1).
Part of this work has been supported by the experimental facilities at Harvard 
MRSEC and the Centre for Nanoscale Systems (CNS), a member of the 
National Nanotechnology Infrastructure Network (NNIN), which is supported 
by the National Science Foundation (ECS-0335765).



\begin{thebibliography}{99}

\bibitem{Callen_book} H. B. Callen, {\it Thermodynamics and an Introduction to 
Thermostatistics} (John Wiley \& Sons, New York, 1985).

\bibitem{heyraud_metois_1983} J. C. Heyraud and J. J. Metois, Surf. Sci., {\bf 128}, 334 (1983).

\bibitem{lowen_ss_1990} H. L\"owen, {\bf 234}, 315 (1990).

\bibitem{Ilett1995} S. M. Ilett, A. Orrock, W. C. K. Poon, and P. N. Pusey, Phys. Rev. E {\bf 51}, 1344 (1995).

\bibitem{poonPRL1999} W. C. K. Poon, F. Renth, R. M. L. Evans, D. J. Fairhurst, 
M. E. Cates, and P. N. Pusey, Phys. Rev. Lett. {\bf 83}, 1239 (1999).

\bibitem{Lekkerkerker2002} V. J. Anderson and H. N. W. Lekkerkerker, Nature {\bf 416}, 811 (2002).

\bibitem{royalJCP2006} C. P. Royall, M. E. Leunissen, A.-P. Hynninen, M. Dijkstra, 
and A. van Blaaderen, J. Chem. Phys. {\bf 124}, 244706 (2006).

\bibitem{ILMR} A. Ivlev, H. L\"owen, G. Morfill, and C. P. Royall, {\it Complex 
Plasmas and Colloidal Dispersions: Particle-Resolved Studies of Classical 
Liquids and Solids} (World Scientific, Singapore, 2012).

\bibitem{Sirota1989} E. B. Sirota, H. D. Ou-Yang, S. K. Sinha, P. M. Chaikin, 
J. D. Axe, and Y. Fujii, Phys. Rev. Lett. {\bf 62}, 1524 (1989).

\bibitem{Monovoukas1989} Y. Monovoukas and A. P. Gast, J. Coll. Interface Sci. {\bf 128}, 533 (1989).

\bibitem{Hynninen2003} A.-P. Hynninen and M. Dijkstra, Phys. Rev. E {\bf 68}, 021407 (2003).

\bibitem{Dupont1993} G. Dupont, S. Moulinasse, J. P. Ryckaert, and M. Baus, Mol. Phys. {\bf 79}, 453 (1993).

\bibitem{Hamaguchi1997} S. Hamaguchi, R. T. Farouki, and D. H. E. Dubin, Phys. Rev. E {\bf 56}, 4671 (1997).

\bibitem{Meijer1991} E. J. Meijer and D. Frenkel, J. Chem. Phys. {\bf 94}, 2269 (1991).

\bibitem{Stevens1993} M. J. Stevens and M. O. Robbins, J. Chem. Phys. {\bf 98}, 
2319 (1993).

\bibitem{yodhPRL2010} Y. Peng, Z.-R. Wang, A. M. Alsayed, A. G. Yodh, and 
Y. Han, Phys. Rev. Lett. {\bf 104}, 205703 (2010).

\bibitem{Watzlawek1999} M. Watzlawek,  C. N. Likos, and H. L{\"o}wen, Phys. Rev. Lett. {\bf 82}, 5289 (1999).

\bibitem{Gottwald2004} D. Gottwald, C. N. Likos, G. Kahl, and H. L{\"o}wen, Phys. Rev. Lett. {\bf 92}, 068301 (2004). 

\bibitem{thomas1996} H. M. Thomas and G. E. Morfill, Nature (London) {\bf 379}, 806 (1996).

\bibitem{morfillRMP} G. E. Morfill and A. V. Ivlev, Rev. Mod. Phys. {\bf 81}, 1353 (2009).

\bibitem{Chu1994} J. H. Chu and Lin I, Phys. Rev. Lett. {\bf 72}, 4009 (1994).

\bibitem{fortov2005}V. E. Fortov, A. V. Ivlev, S. A. Khrapak, A. G. Khrapak, and 
G. E. Morfill, Phys. Rep. {\bf 421}, 1 (2005).

\bibitem{vaulina2000} O.S. Vaulina, and S.A. Khrapak, J. Exp. Theor. Phys. {\bf 90}, 287 (2000).

\bibitem{yethiraj_nature} A. Yethiraj and A. van Blaaderen, Nature, {\bf 421}, 513 (2003).

\bibitem{yethiraj.nature} A. Yethiraj, Soft Matter {\bf 3}, 1099 (2007).

\bibitem{Gasser:2001aa} U. Gasser, E. R. Weeks, A. Schofield, P. N. Pusey, 
and D. A. Weitz, Science {\bf 292}, 258 (2001).

\bibitem{palbergJCP2005} P. Wette, H. J. Sch\"ope, and T. Palberg, J. Chem. Phys. {\bf 123} , 174902 (2005).

\bibitem{tanaka2010PNAS} T. Kawasaki and H. Tanaka, Proc. Natl. Acad. Sci. USA {\bf 107}, 14036 (2010).

\bibitem{Wang2012} Z. R. Wang, F. Wang, Y. Peng, Z. Y. Zheng, and Y. L. Han, Science {\bf 338}, 87 (2012).

\bibitem{Poon2009} E. Zaccarelli, C. Valeriani, E. Sanz, W. C. K. Poon, M. Cates, 
and P. N. Pusey, Phys. Rev. Lett. {\bf 103}, 135704 (2009).

\bibitem{Zhou2011} H. Zhou, S. Xu, Z. Sun, X. Du, and L. Liu, Langmuir {\bf 27}, 7439 (2011).

\bibitem{Shabalin2016} A. G. Shabalin, J.-M. Meijer, R. Dronyak, O. M. Yefanov, 
A. Singer, R. P. Kurta, U.Lorenz, O. Y. Gorobtsov, D. Dzhigaev, S. Kalbfleisch, 
J. Gulden, A. V. Zozulya, M. Sprung, A. V. Petukhov, and I. A. Vartanyants, 
Phys. Rev. Lett. {\bf 117}, 138002 (2016).

\bibitem{oettelPRL} T. Schilling, H. J. Sch\"ope, M. Oettel, G. Opletal, and 
I. Snook, Phys. Rev. Lett. {\bf 105}, 025701 (2010).

\bibitem{allahyarov2015} E. Allahyarov, K. Sandomirski, S. U. Egelhaaf, and 
H. L\"owen, Nat. Comm. {\bf 6}, 7110 (2015).

\bibitem{palbergJCP} N. Lorenz, and T. Palberg, J. Chem. Phys. {\bf 133}, 104501 (2010).

\bibitem{Sprakel2017} J. Sprakel, A. Zaccone, F. Spaepen, P. Schall, and 
D. A. Weitz, Phys. Rev. Lett. {\bf 118}, 088003 (2017).

\bibitem{Loehr2016} J. Loehr, A. Ortiz-Ambriz, and P. Tierno, Phys. Rev. Lett. {\bf 117}, 168001 (2016).

\bibitem{Weeks:2000aa} E. R. Weeks, J. C. Crocker, A. C. Levitt, A. Schofield, 
and D. A. Weitz, Science {\bf 287}, 627 (2000).

\bibitem{tanaka2010NM} H. Tanaka, T. Kawasaki, H. Shintani, and K. Watanabe, Nat. Mat. {\bf 9}, 324 (2010).

\bibitem{tanaka2012NC} M. Leocmach and H. Tanaka, Nat. Comm. {\bf 3}, 974 (2012).

\bibitem{Zhang2016} B. Zhang and X. Cheng, Phys. Rev. Lett. {\bf 116}, 098302 (2016).

\bibitem{Weeks2009} J. Hernandez-Guzman and E. Weeks, Proc. Natl. Acad. Sci. USA {\bf 106}, 
15198 (2009).

\bibitem{Ramsteiner2010} I. B. Ramsteiner, D. A. Weitz, F. Spaepen, Phys. Rev. E {\bf 82}, 041603 (2010).

\bibitem{Schall2011} V. D. Nguyen, M. T. Dang, B. Weber, Z. Hu, and P. Schall, Adv. Mater. {\bf 23}, 2716 (2011).

\bibitem{Palberg2016} T. Palberg, P. Wette, and D. M. Herlach, Phys. Rev. E {\bf 93}, 022601 (2016).

\bibitem{Heinonen2013} V. Heinonen, A. Mijailovic, C. V. Achim, T. Ala-Nissila, 
R. E. Rozas, J. Horbach, and H. L\"owen, J. Chem. Phys. {\bf 138}, 044705 (2013).

\bibitem{yethiraj.PRL2004} A. Yethiraj, A. Wouterse, B. Groh, and A. van 
Blaaderen, Phys. Rev. Lett. {\bf 92}, 058301 (2004).

\bibitem{crocker2012} M. T. Casey, R. T. Scarlett, W. B. Rogers, I. Jenkins, 
T. Sinno, and J. C. Crocker, Nat. Comm. {\bf 3}, 1209 (2012).

\bibitem{yodhNat.mat.2015} Y. Peng, F. Wang, Z. Wang, A. M. Alsayed, 
Z. Zhang, A. G. Yodh, and Y. Han, Nat. Mat. {\bf 14}, 101 (2015).

\bibitem{Maire2016} E. Maire, E. Redston, M. Persson Gulda, D. A. Weitz, 
and F. Spaepen, Phys. Rev. E {\bf 94}, 042604 (2016).

\bibitem{Evers2013} F. Evers, R. D. L. Hanes, C. Zunke, R. F. Capellmann, 
J. Bewerunge, C. Dalle-Ferrier, M. C. Jenkins, I. Ladadwa, A. Heuer, 
R. Castaneda-Priego and S. U. Egelhaaf, Eur. Phys. J. Special 
Topics {\bf 222}, 2995 (2013).

\bibitem{Laurati2017} M. Laurati, P. Ma{\ss}hoff, K. J. Mutch, S. U. Egelhaaf, 
and A. Zaccone, Phys. Rev. Lett. {\bf 118}, 018002 (2017).

\bibitem{Sentjabrskaja2015} T. Sentjabrskaja, P. Chaudhuri, M. Hermes, 
W. C. K. Poon, J. Horbach, S. U. Egelhaaf, and M. Laurati, Sci. Rep. {\bf 5}, 11884 (2015).

\bibitem{Schall2007} P. Schall, D. A. Weitz, and F. Spaepen, Science {\bf 318}, 1895 (2007).

\bibitem{Hsu_2005} F. Hsu, E. R. Dufresne, and D. A. Weitz, Langmuir {\bf 21}, 4881 (2005).

\bibitem{Mukherjee1993} K. Mukherjee, S. P. Mouli, and D. C. Mukherjee, Langmuir {\bf 9}, 1727 (1993).

\bibitem{Kanai2015} T. Kanai, N. Boon, P. J. Lu, E. Sloutskin, A. B. Schofield, 
F. Smallenburg, R. van Roij, M. Dijkstra, and D. A. Weitz, Phys. 
Rev. E {\bf 91}, 030301 (2015).

\bibitem{Zulauf1979} M. Zulauf and H.-F. Eicke, J. Phys. Chem. {\bf 83}, 480 (1979).

\bibitem{VO} E. J. W. Verwey and J. T. G. Overbeek, {\it Theory of the Stability of 
Lyophobic Colloids} (Elsevier, Amsterdam, 1948).

\bibitem{Bocquet2002} L. Bocquet, E. Trizac, and M. Aubouy, J. Chem. Phys. {\bf 117}, 8138 (2002).

\bibitem{foot} The samples are filled in glass capillaries with a size of about 
$15 \times 5 \times 0.2$~mm$^3$ 
and left for two days to allow for equilibration before volumes of 
$145 \times 145 \times 60\;\mu$m$^3$, corresponding to 
$512 \times 512 \times 478$ voxels, and located $20\,\mu$m from the capillary 
wall are imaged using a confocal microscope (Leica TCS SP5 with a $40\times$ 
oil-immersion objective).

\bibitem{Hamaguchi1996} S. Hamaguchi, R. T. Farouki, and D. H. E. Dubin, 
J. Chem. Phys. {\bf 105}, 7641 (1996).

\bibitem{Kremer1986} K. Kremer, M. O. Robbins, and G. S. Grest,
Phys. Rev. Lett. {\bf 57}, 2694 (1986).

\bibitem{Sengupta1991} S. Sengupta and A. K. Sood, Phys. Rev. A {\bf 44}, 
1233 (1991).

\bibitem{Smallenburg2011} F. Smallenburg, N. Boon, M. Kater, M. Dijkstra, and 
R. van Roij, J. Chem. Phys. {\bf 134}, 074505 (2011).

\bibitem{Schoepe1998} H. J. Sch{\"o}pe, T. Decker, and T. Palberg, J. Chem. Phys. {\bf 109}, 10068 (1998).

\bibitem{deAnda2015} I. R. de Anda, A. Statt, F. Turci, and C. P. Royall, Contrib. Plasma Phys. {\bf 55}, 172 (2015).

\bibitem{rogers} R. B. Rogers and B. J. Ackerson, Philos. Mag. {\bf 91}, 682 (2011).

\bibitem{crocker.grier.1996} J. C. Crocker and D. G. Grier, J. Colloid Interface Sci. {\bf 179}, 298 (1996).

\bibitem{frenkel1996} P. R. ten Wolde, M. J. Ruiz-Montero and D. Frenkel, Faraday Disc. {\bf 104}, 93 (1996).

\bibitem{tenWolde1995} P. R. ten Wolde, M. J. Ruiz-Montero, and D. Frenkel, Phys. Rev. Lett. {\bf 75}, 2714 (1995).

\bibitem{Mullins1957} W. W. Mullins, {\it Theory of thermal grooving}, J. Appl. Phys. {\bf 28}, 333 (1957).

\bibitem{ratke} H. J. Vogel and L. Ratke, Acta. Metal. Mater. {\bf 39}, 641 (1991).

\bibitem{lin2016} S.-C. Lin, M.-W. Liu, M. P. Gururajan, and K.-A. Wu, Acta Materialia {\bf 102}, 364 (2016).

\bibitem{Haertel2012} A. H{\"a}rtel, M. Oettel, R. E. Rozas, S. U. Egelhaaf, J. Horbach, 
and H. L{\"o}wen, Phys. Rev. Lett {\bf 108}, 226101 (2012).

\bibitem{Hillert} M. Hillert {\it Phase Equilibria, Phase Diagrams and Phase Transformations: Their Thermodynamic Basis} (Cambridge Universitxy Press, Cambridge, 1998).

\bibitem{Toth2010} G. I. T{\'o}th, G. Tegze, T. Pusztai, G. T{\'o}th, and L. Gr{\'a}n{\'a}sy, J. Phys.: Condens. Matter {\bf 22}, 364101 (2010).

\bibitem{Emmerich2007} H. Emmerich, Philos. Mag. Lett. {\bf 87}, 795 (2007).

\bibitem{Sosso2016} G. C. Sosso, J. Chen, S. J. Cox, M. Fitzner, P. Pedevilla, A. Zen, and A. Michaelides, Chem. Rev. {\bf 116}, 7078 (2016).

\bibitem{Dhont1996} J. K. G. Dhont, Phys. Rev. Lett. {\bf 76}, 4269 (1996).

\bibitem{Lettinga2004} M. P. Lettinga, H. Wang, and J. K. G. Dhont, Phys. Rev. E {\bf 70}, 061405 (2004).

\bibitem{Bechinger2001} C. Bechinger, M. Brunner, and P. Leiderer, Phys. Rev. Lett. {\bf 86}, 930 (2001).





\end{thebibliography}
\end{document}